\documentclass{elsart}

 \usepackage{epsfig}

\usepackage{amssymb}

\begin{document}

\begin{frontmatter}



\title{
Universal local versus unified global scaling laws
in the statistics of seismicity
}


\author{\'Alvaro  Corral}
\ead{Alvaro.Corral@uab.es}
\address{
Departament de F\'\i sica,
Facultat de Ci\`encies,
Universitat~Aut\`onoma~de~Barcelona,
E-08193 Bellaterra,
Spain
}

\begin{abstract}
The unified scaling law for earthquakes,
proposed by Bak, Christensen, Danon and Scanlon,
is shown to hold worldwide, as well as for areas
as diverse as Japan, New Zealand, Spain or New Madrid.
The scaling functions that account for the 
rescaled recurrence-time probability densities
show a power-law behavior for long times,
with a universal exponent about (minus) 2.2.
Another decreasing power law governs short times, 
but with an exponent that may change from one area
to another.
This is in contrast with a spatially independent,
time-homogenized version of Bak {\it et al.}'s
procedure, which seems to present a universal
scaling behavior.
\end{abstract}

\end{frontmatter}

\section{Introduction}

Everybody will agree that earthquakes are a complex phenomenon.
Indeed, one single event can generate a bunch of research papers
(several of them published even in {\it Nature} or {\it Science}),
each telling a part of the story of the quake. 
Essentially, these articles would argue that
if the tectonic forces were like this and that 
and after the effect of diverse factors 
and some calculations, we would end with precisely that particular earthquake.
This research provides important information for specific mechanisms
triggering earthquakes, ``one explanation for each earthquake''.

However, another point of view is possible.
As an alternative to the described reductionism, 
Bak was claiming the necessity of a general theory encompassing
all earthquakes \cite{Bak96}, from the very small (imperceptible by humans)
to the largest, catastrophic ones (``killing hundreds of thousands of people''),
irrespective of their location at the boundaries or interior of the plates, 
depth, and other tectonic details.
A key clue signaling 
the unity 
of the phenomenon
is the existence of the Gutenberg-Richter law 
\cite{Kagan94,Turcotte97}, 
which states that 
(for any region and for a large enough period of time) the number of earthquakes
decreases exponentially as a function of their magnitude.
If the seismic processes did not form a whole, 
how could it be that all that variety of events 
conspire together to align onto such a simple curve?

The first step towards a theory of earthquakes should consist
on identifying what kind of dynamical process
we are dealing with \cite{Bak_debate}: 
``is it periodic? Is it chaotic? Is it random in space and time?''
From our present knowledge, 
the best candidate 
is self-organized criticality
(SOC) \cite{Bak96,Jensen,Turcotte99,Hergarten_book}.
The analogies between earthquakes and SOC systems are clear
\cite{Tang,Sornette89,Ito90}:
the Earth crust accumulates energy 
(supplied by slow convective motion in the mantle) 
in the form of elastic deformation
at a very slow rate.
At some point the stress
cannot be sustained an a rupture initiates,
propagating very fast through a fault by means of a domino effect, 
giving rise to an earthquake.
So, we have the basic ingredients for SOC, i.e., 
a long-term balance between slow driving and
fast avalanches in a spatially extended system consisting
on many interacting parts.
The Gutenberg-Richter law is again crucial in this picture,
as it implies that there is not a characteristic scale for 
the energy dissipated during an earthquake 
(this is so because the energy increases exponentially with
the magnitude [about a factor 30 in the energy for each unit
in the magnitude], and therefore, the exponential frequency-magnitude
relation transforms into a power-law distribution of dissipated
energies \cite{Turcotte97}, which is the indication of scale invariance,
i.e., criticality).
Therefore, one may talk about the crust as being at a critical
state, but in contrast to equilibrium critical points,
this state has to arise spontaneously, as an attractor of the dynamics.

In addition to the Gutenberg-Richter law, there are other indicators
of criticality or scale invariance in seismicity, 
as the fractal distribution of hypocenters
or epicenters \cite{Kagan94,Turcotte97,Kagan80}, 
and the Omori law, which tells us that the decay 
of the rate of seismic activity after a large event 
does not present any characteristic time \cite{Utsu}. 
Further, the structure of faults and tectonic plates is also fractal
\cite{Turcotte97,Okubo,plates}.
For these reasons, although the SOC paradigm is represented by
a sandpile \cite{Bak87}, earthquakes may be considered as the
clearest illustration of how a real SOC system would look like
\cite{Bak96}.

Despite the initial opposition of the very conservative geophysics
community to the idea of earthquakes as a SOC phenomenon \cite{Bak96},
nowadays SOC is very seriously considered by many professional
seismologists. 
It seems that scientific evolution in the solid-Earth sciences
takes place mostly after great painful controversies \cite{Hallam};
and earthquakes are a field where the debate is open
\cite{debate}.

%

A great hallmark in the earthquake-as-a-SOC-phenomenon
development was the release of the Olami,
Feder, and Christensen (OFC) model \cite{Olami}
(for a summary of
the rich dynamical behavior of its variations see 
Ref. \cite{Perez}).
However, it is a subsequent surprising proposal by Ito 
what calls our interest here \cite{Ito95}:
the well-studied Bak-Sneppen (BS) model \cite{Bak93}, introduced to account
for biological evolution, reproduced some other properties
of earthquakes. Ito measured for a California catalog the same quantities
used to characterize the BS model \cite{Paczuski96},
in particular, the first-return-time distribution:
this is the probability that the activity returns
(for the first time) at a given spatial location after a certain
time.
As the locations of earthquake occurrence are continuous
(in contrast to the BS model),
Ito divided the area covered by the catalog 
into small regions of $1^\circ $ latitude $\times 1^\circ $
longitude and measured the return times to these regions.
(Another important difference between the model and
real earthquake occurrence is that the former is spatially homogeneous,
whereas earthquake epicenters draw a fractal over the
Earth surface.)
The results seemed compatible with a power-law distribution
(as in the BS model), but clearly, a more in-depth investigation
was needed.

It was Bak, together with Christensen, Danon and Scanlon,
who re-opened the problem 
\cite{Bak02,Christensen02}.
In essence, they used Ito's procedure with the addition of a lower
bound $M_c$ for the magnitude, in such a way that events
with magnitude $M$ below $M_c$ were disregarded.
This is necessary to 
avoid spatial and time variations 
in the completeness of the catalogs and to
ensure that 
no events (or not many) with $M \ge M_c$ 
are missing.
A power-law first-return-time distribution
(followed by a faster decay) was indeed found,
but the exponent was different from Ito's one
(the faster decay was later identified as
another power law \cite{Corral03}).
However, Bak {\it et al.} 
also introduced a crucial element, which was 
the study of the distribution under the variation 
of the two parameters of the procedure:
the lower bound $M_c$ and
the size $L$ of the small regions.
Remarkably, a scaling analysis showed that different distributions
corresponding to different values of $L$ and $M_c$ collapsed
onto a single curve under rescaling of the axes by a factor
$10^{b M_c}/L^{d_f}$, where the numerator comes from the 
Gutenberg-Richter relation and the denominator accounts for
the fractal distribution of epicenters.
In this way, it is appropriate to talk about a 
{\it unified scaling law for earthquakes}.


The law is a direct consequence of Bak's philosophy
applied to earthquakes, which can be summarized by:
%
1)
Don't care about the tectonic environment.
(The small regions in which California is divided 
are independent of it, in contrast with traditional studies.)
%
%
2)
Don't care about aftershocks, foreshocks, or mainshocks
(none of these events are removed,
all are equally treated,
again at variance with usual approaches).
After all, there is nothing in the seismograms that
differentiates these events;
it is only from their relation with the other events that
these categories can be established,
and not without unambiguities.
%
%
3)
Don't care about temporal heterogeneity.
%

We are going to explore in detail Bak {\it et al.}'s
unified scaling law; first, introducing some variations
to their procedure, and next, turning to their method
and extending it {\it everywhere}.
Needless to mention, the original flow of ideas 
is the opposite as the one presented here; 
nevertheless, we believe it is more direct to start
with our perspective first (at least for us).


\section{Universal scaling law for 
local distributions of recurrence times}

Let us consider spatial regions of arbitrary shape, 
which can be of small size (as in Ito's paper \cite{Ito95})
but also large.
In contrast to Ito and Bak {\it et al.} \cite{Bak02,Christensen02}, 
we concentrate 
only on one of these regions, where (in the same way as
Bak {\it et al.}) earthquakes 
with magnitude $M$ above a lower bound $M_c$ are selected.
Then, if $t_i$ denotes the time of occurrence of the 
$i-$th earthquake in the spatial and magnitude windows considered,
we calculate the {\it recurrence time} between events $i$ and $i-1$ as
$
\tau_i=t_i-t_{i-1}.
$
This time is the same first-return time defined by Ito,
or the waiting time in the language of Bak {\it et al.},
or interoccurrence or inter-event time in other papers.

The probability density of the recurrence times can easily be obtained;
however, in order to pay attention to all the time scales
involved in the process it will be convenient to look at time
in logarithmic scale.
The first data set to start these measurements is a global
earthquake catalog, as the rate of seismic occurrence there
(defined as number of earthquakes per unit time)
is fairly constant, and therefore stationary.
For instance, for the NEIC worldwide catalog \cite{Corral04}, 
covering a period from 1973 to 2002 (included),
and for several very large regions
we get the results displayed 
in Fig. 1 (top curve),
after rescaling the axes by the seismic rate.
All the distributions lie onto a single curve, 
so,
\begin{equation}
D_{xy}(\tau) = R_{xy} f(R_{xy} \tau ), 
\end{equation}
where, for a region of spatial coordinates $xy$,
$D_{xy}(\tau)$ is the probability density
that the recurrence time is around a value $\tau$ 
and $R_{xy}$ is the mean rate of seismic occurrence
(or activity). 
It is implicit that both $D_{xy}(\tau)$ and $R_{xy}$ 
depend as well on $M_c$ and the size of the region; to be concrete,
if the region is kept fixed the rate depends exponentially on
the magnitude, following the Gutenberg-Richter relation:
$R_{xy}=N_{xy} 10^{-b M_c}$,
with $N_{xy}$ the (hypothetical) number of events
per unit time
in the region with magnitude above 0.

\begin{figure}
\centering
\includegraphics[height=6cm]{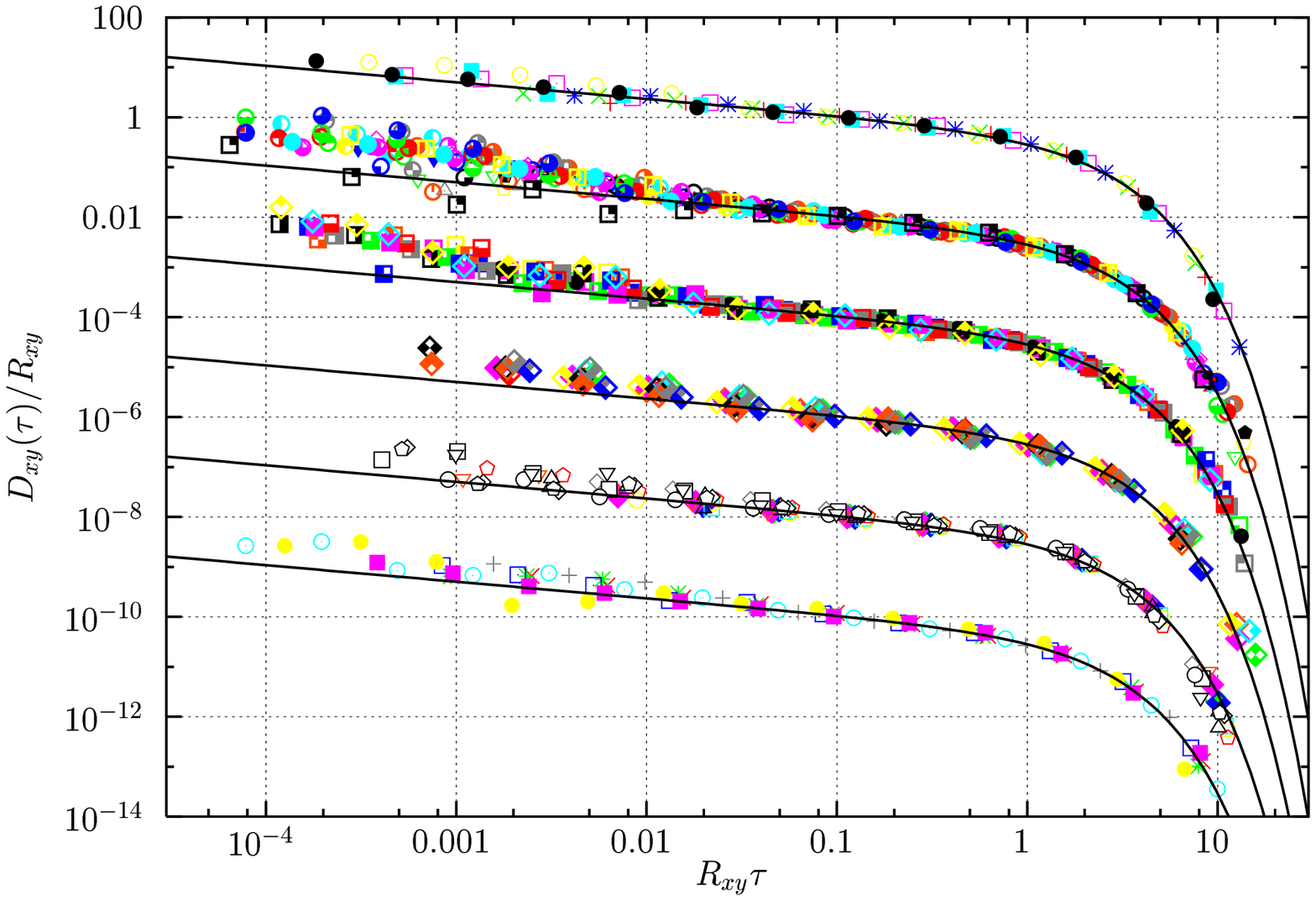}
\caption{
Single-region recurrence-time probability densities, 
$D_{xy}$, after rescaling with the rate $R_{xy}$.
The five data sets correspond, from top to bottom,
to: 
1, the NEIC worldwide catalog for regions with
$L \ge 180^\circ$, 1973 to 2002;
2, NEIC with $L \le 90^\circ$ (same period of time);
3, Southern California, 1984 to 2001, 1988 to 1991, and 1995 to 1998;
4, Northern California, 1998 to 2002,
5, Japan, 1995 to 1998, and New Zealand, 1996 to 2001;
6 (bottom), Spain, 1993 to 1997, New Madrid, 1975 to 2002, and Great Britain,
1991 to 2001.
The distribution sets have been multiplied by 
$10^{0}, 10^{-2}, 10^{-4}, 10^{-6}$, $10^{-8}$, and $10^{-10}$,
for clarity sake.
A total of 82 different distributions are shown,
with the size of the regions from $0.16^\circ$
to the whole world, and $M_c$ from 1.5 to 7.5.
Recurrence times go from 2 min to about 1.5 years;
values of $\tau < 2$ min are not shown.
The lines correspond to $f(\theta)$.
}       
\end{figure}

The scaling function $f$ can be represented by 
\begin{equation}
f(\theta)=\frac{C |\delta|}{a \Gamma(\gamma/\delta)}
\left( \frac \theta a \right)^{\gamma-1}e^{-(\theta/a)^\delta},
\end{equation}
which has indeed a very general shape.
If $\gamma$ and $\delta$ are positive, the former controls the shape
for small $\theta$ and $\delta$ the shape at large $\theta$;
the situation is reversed if both parameters are negative;
$a$ is a scale parameter and $C$ a normalization correction. 

If regions of smaller size are considered,
the rate turns nonstationary,
giving rise to heterogeneities in time.
This is due to large
earthquakes, which provoke a kind of ``avalanches
of earthquakes'', i.e., the aftershock sequences.
In order to compare with the worldwide case, 
we concentrate in space-time windows in which
the rate keeps stationary, in other words,
we stay away of time periods which include
very prominent aftershock sequences (by now),
in opposition to Bak {\it et al.}
(The simplest way to recognize stationarity
is by a linear increase with time of the accumulated number
of earthquakes in a region.)
Using the NEIC data and several regional and local
catalogs (Southern California, Japan, Spain, Great Britain \cite{Corral04},
New Zealand, New Madrid \cite{Corral04b}, 
and also Northern California \cite{NC}), 
we find the results displayed in Fig. 1,
taking regions of $L$ degrees in longitude and
$L$ degrees in latitude.
The behavior of the distributions is identical to the previous case,
collapsing under rescaling onto the same universal curve $f$.
(The deviations at short times are due to a nonstationary rate
at this time scale, provoked by small aftershock sequences.)


A massive least-square fit using many regions from the NEIC catalog
and several values of $M_c$ and $L$ yields \cite{Corral04}
$\gamma= 0.67 \pm 0.05$, $\delta  = 1.05 \pm 0.05$
and $a = 1.64 \pm 0.15$;
so, $\delta$ can be considered to be one and we have
a decreasing power law with exponent $1-\gamma$ about 0.3
accelerated by an exponential term at large times.
This type of distribution indicates that earthquakes
cluster in time, 
not only for sequences of aftershocks
(as it is well known)
but even for ``background seismicity''.
The counterintuitive consequences of this phenomenon
for the time evolution of seismic hazard
are analyzed in Ref. \cite{Corral04b}.

Finally, the scaling law we propose is valid beyond the
stationary limit, replacing the mean rate $R_{xy}$
by its instantaneous value, $r_{xy}(t)$.
In this way, the probability density $\Psi$ of 
$\theta \equiv r_{xy}(t)\tau$ satisfies
$\Psi = f(r_{xy}(t) \tau)$,
for aftershock sequences where the rate decays
following the Omori law, $r_{xy}(t)=A_{xy}/t^p$,
with the origin of time at the mainshock.
Remarkably, the scaling function $f$ turns out to be 
the same as in the stationary case \cite{Corral04}.

As all the data analyzed are well fit by the same function
$f$, with the same values of the parameters, 
we may talk about a universal scaling law.
Nevertheless, this law cannot be designated as 
unified, since the scaling only includes the
Gutenberg-Richter law, but not
the fractal dimension of the epicenters,
in contrast to Bak {\it et al.}'s law.
And in terms of its definition in space,
$D_{xy}$ may be referred to as a local distribution
(although the size of the $xy-$region
can grow to reach the total area covered 
by the catalog).

\section{Unified scaling law beyond Southern California}

Now it is time to pay attention to the original approach
of Bak {\it et al.}
We only have to consider all the (non-overlapping) regions of size $L$ 
necessary to cover completely
a much larger area (Southern California, Japan, 
the whole world, or New Madrid)
and expand the time windows without bothering
about the nonstationarities of the seismic rate.
We measure for each $xy-$region the recurrence times
$\tau_i^{(xy)}$ in the same way as before,
with the difference that all these series of recurrence
times are counted into one single probability density, 
${\mathcal{D}}(\tau)$,
which we may call global 
(in contrast with our local version).
It is found that ${\mathcal{D}}(\tau)$ scales,
under the change of $L$ and the lower bound $M_c$, 
as:
\begin{equation}
{\mathcal{D}}(\tau)={\mathcal{R}}F({\mathcal{R}} \tau),
\end{equation}
with ${\mathcal{R}}$ the mean value of the local
mean rate $R_{xy}$, calculated over all the $xy-$regions
of size $L$ with seismic activity, i.e.,
${\mathcal{R}}={\sum_{xy} R_{xy}}/{n}$,
where ${n}$ is the number of such regions.
From the equation for $R_{xy}$ 
and the scaling of ${n}$ with $L$,
${n}=({\ell}/L)^{d_f}$, we get,
${\mathcal{R}}=N (L/{\ell})^{d_f} 10^{-b M_c}$,
with $N=\sum_{xy} N_{xy}$ and ${\ell}$ a rough measure
of the linear size of the total area under study
(in degrees).

\begin{figure}
\centering
\includegraphics[height=6cm]{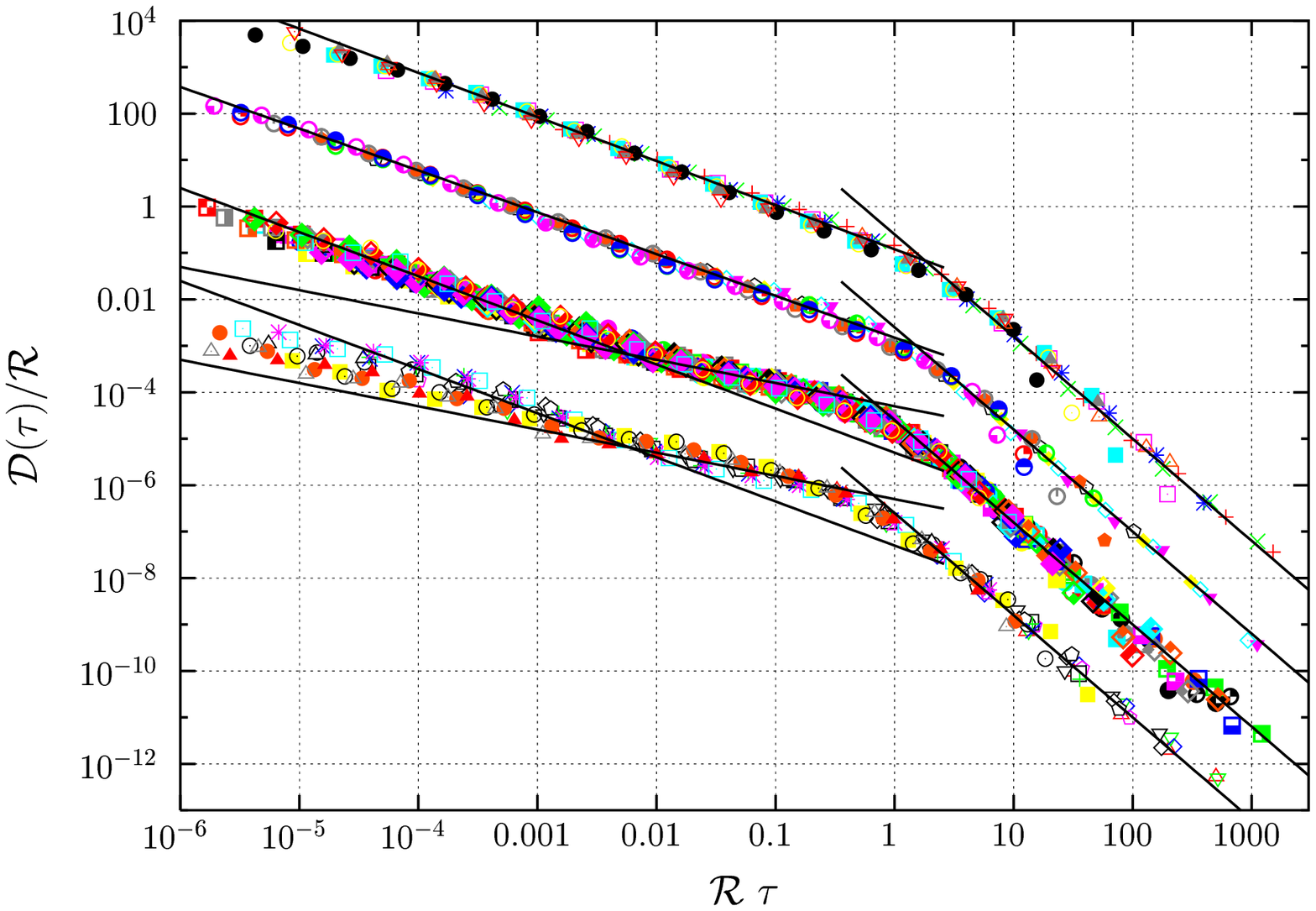}
\caption{
Recurrence-time probability densities,
calculated following Bak {\it et al.}'s
procedure, after rescaling by $\mathcal{R}$.
84 distributions are shown, with $L$ 
ranging from $0.039^\circ$ to $45^\circ$,
and $1.5 \le M_c \le 6$.
The curves are shifted by factors
$10^{0}, 10^{-2}, 10^{-4}$ and 
$10^{-6}$, and correspond to:
1 (top), Southern California, 1984 to 2001;
2, Northern California, 1985 to 2003;
3, Southern California, 1988 to 1991 (stationary rate), 
NEIC, 1973 to 2002, Japan, 1995 to 1998, and Spain, 1993 to 1997;
4 (bottom), New Zealand, 1996 to 2001, and New Madrid, 1975 to 2002.
Short or intermediate times are fit by:
1 (top), $0.12/\theta^{0.95}$; 2, $0.15/\theta^{0.9}$,
3 and 4 (bottom), $0.05/\theta^{0.95}$ and $0.5/\theta^{0.5}$,
with $\theta \equiv {\mathcal{R}}\tau$.
In all cases the long-time tail is fit by 
$0.25/\theta^{2.2}$.
The times in the horizontal axis span from
2 min to about 20 years.
Recurrence times smaller than 2 minutes are not shown,
except for Japan and the NEIC catalog where the minimum times
are 4 min and 10 min, respectively.
}       
\end{figure}

In addition to Southern California \cite{Bak02,Christensen02}, 
the scaling relation for ${\mathcal{D}}$ is valid
for the catalogs studied in the previous section, see Fig. 2.
However, the scaling function $F$ does not seem to be universal;
the clear decreasing two-power-law behavior of Southern California
\cite{Corral03},
with exponent for small $\tau$ about 0.95 
and 2.2 for large $\tau$, 
changes slightly for Northern California
(exponents 0.9 and 2.2), 
but becomes more complicated for the other catalogs,
with the appearance of an intermediate bump.
Figure 2 shows how this bump can be approximated 
by another decreasing power law,
with exponent 0.5, roughly.
Further, for New Madrid  
there is no trace of the 0.95 exponent,
and for New Zealand the behavior is not clear.
The only exponent that seems to be universal
is the one for large times, 2.2 in all cases.
The unified scaling law could also hold for Great Britain,
but the few data considered there, only about 500 events,
makes the statistics too poor for small and
intermediate times.




Let us mention that the probability densities of
Bak {\it et al.} and ours,
in addition to quite different shapes, 
have different meanings too.
In our case, we provide the probability of return of
an earthquake 
for a given $xy-$region
of size $L$, with the only information required
being the mean seismic rate there, $R_{xy}$.
On the other hand, Bak {\it et al.}'s distribution
gives the return probability if one does not know 
in which region of size $L$ of a much larger area
(like Southern California) one is
(or you know the region but you don't have 
knowledge about the rate $R_{xy}$),
and it is necessary to know the average rate
${\mathcal{R}}$ for regions of size $L$.
The relation between these two approaches are studied 
in Ref. \cite{Corral03}.

As a final conclusion, we see that scaling is an
intrinsic characteristic of seismicity.
Both the unified scaling law and our local
approach show that the distributions of recurrence
times scale with some average value of the rate of
seismic activity.
This is a clear consequence of the scale-invariant structure
of seismic occurrence in time, space and magnitude.
Whatever triggers earthquakes 
operates in the same way
at all spatial and temporal scales.

\section{Acknowledgement} 

The author would have been unable to undertake this research
without the seminal contribution of Per Bak, whose personal warmth
was also very much appreciated.
M. Bogu\~n\'a, K. Christensen, and the Ram\'on y Cajal program
have been important at different stages of this process.

\end{document}